# AdaBoost-Based Efficient Channel Estimation and Data Detection in One-Bit Massive MIMO

Majdoddin Esfandiari, *Student Member, IEEE*, Sergiy A. Vorobyov, *Fellow, IEEE*, and Robert W. Heath Jr., *Fellow, IEEE*

*Abstract*—The use of one-bit analog-to-digital converter (ADC) has been considered as a viable alternative to high resolution counterparts in realizing and commercializing massive multiple-input multiple-output (MIMO) systems. However, the issue of discarding the amplitude information by one-bit quantizers has to be compensated. Thus, carefully tailored methods need to be developed for one-bit channel estimation and data detection as the conventional ones cannot be used. To address these issues, the problems of one-bit channel estimation and data detection for MIMO orthogonal frequency division multiplexing (OFDM) system that operates over uncorrelated frequency selective channels are investigated here. We first develop channel estimators that exploit Gaussian discriminant analysis (GDA) classifier and approximated versions of it as the so-called weak classifiers in an adaptive boosting (AdaBoost) approach. Particularly, the combination of the approximated GDA classifiers with AdaBoost offers the benefit of scalability with the linear order of computations, which is critical in massive MIMO-OFDM systems. We then take advantage of the same idea for proposing the data detectors. Numerical results validate the efficiency of the proposed channel estimators and data detectors compared to other methods. They show comparable/better performance to that of the state-of-the-art methods, but require dramatically lower computational complexities and run times.

## I. INTRODUCTION

Utilization of a large number of antennas at the base station (BS) in communication systems have been explored for the purpose of enhancing data rates and network capacity [1], [2]. Massive multiple-input multiple-output (MIMO) communication systems have been demonstrated to offer remarkable advantages, but the hardware cost and high power consumption are two main difficulties (among others), hindering their commercial usage. To address these issues, first analog-to-digital converters (ADCs) have been recognized as one of the parts of the receivers that have high power consumption and expensive price [3], [4]. Then, employing low-resolution (ADCs) has been suggested as a viable alternative instead of using high-resolution counterparts [5], [6]. However, the use of low-resolution ADCs in multi-user MIMO-OFDM systems poses several challenges in the receiver design. For instance, the non-linearities caused by few bit quantizers may prohibit us from exploiting conventional receivers like zero-forcing (ZF) and minimum mean square error (MMSE) detectors [7]. The reason is that the conventional procedure of isolating narrowband OFDM subcarries using a discrete Fourier transform (DFT) at the receiver is not valid when low-resolution ADCs are used. Instead, different receiver architectures need to be employed/designed to process the baseband time-domain signals for the tasks such as channel estimation, and/or data detection.

Channel estimation and/or data detection in massive MIMO systems with one-bit ADCs have been explored in several papers, considering the cases of single-carrier (SC) and multi-carriers (MC) signalling. The authors of [8] have revised the non-convex optimization problem of the maximum likelihood (ML) channel estimator and proposed a sub-optimal channel estimator referred to as near-ML (nML). The same methodology has been used to develop the nML-based data detector as well. Convex optimization approaches have been exploited in [9] for estimating MC-OFDM, whereas a data detector has been developed based on a soft-output MMSE algorithm. In [10], the Bussgang decomposition [11] has been employed to develop Bussgang-based minimum mean-squared error (BMMSE) channel estimators and data detectors for both SC and MC-OFDM systems. Analogous to [10], the authors of [12] took advantage of the Bussgang decomposition to estimate the optimal nonzero thresholds in the problem of one-bit quantizer design. Multiple works such as [13]- [17] have considered the problem of joint channel estimation and data detection, where the known pilot sequence is augmented with a portion of detected data to build a longer virtual pilot sequence and subsequently utilize it to refine the channel estimate. For instance, the authors of [13] have developed a bilinear generalized approximate message passing (BiGAMP) method, while the authors of [16] have proposed a variational Bayesian (VB) algorithm to do so.

One interesting idea presented by different researchers is to treat one-bit channel estimation and data detection as binary classification problems, where the output of one-bit ADCs can be viewed as class labels. Moreover, a proper transformation of the known pilots or channel state information (CSI) plays the role of the classification features, while the unknown channel/data vectors act as the corresponding separating hyperplanes. For instance, the binary soft-margin support vector machine (SVM) has been considered by the authors of [17] and [18] as a powerful method to estimate the one-bit channel

M. Esfandiari and S. A. Vorobyov are with Department of Information and Communications Engineering, Aalto University, PO Box 15400, 00076 Aalto, Finland. (Emails: majdoddin.esfandiari@aalto.fi; sergiy.vorobyov@aalto.fi).

R. W. Heath Jr. is with Department of Electrical and Computer Engineering, University of California San Diego, Jacobs Hall, EBU1, CA, US 92093-0021. (Email: rwheathjr@ucsd.edu).

estimation and data detection in SC and MC-OFDM scenarios. Although the soft-margin SVM-based estimators have good properties, their performance relies on careful hyperparamer selection. Deep neural networks (DNN) have been also used for one-bit channel estimation in several works such as [19]-[21]. The main disadvantage of such estimators is that not only a sufficiently large data set is required for the training process, but the offline training procedure needs to be executed carefully. In [22]- [24], several blind/semi-blind learning-based data detectors have been presented for massive MIMO systems that employ one-bit ADCs.

Angular domain channel estimators have been reported in [25]- [32]. In [25] and [26], compressive sensing (CS) techniques have been adopted to recover sparse millimeter wave (mmWave) channels quantized by few-bit ADCs. The authors of [30] have considered the combination of harmonic retrieval methods with a modified expected-maximization GAMP (EM-GAMP) to devise an angular domain one-bit mmWave channel estimation approach called gridless GAMP (GL-GAMP). For such channels, a sparsity enforcing with Toeplitz matrix reconstruction (SE-TMR) method was also presented in [31] recently. Moreover, the authors of [32] have used the Toeplotz matrix reconstruction notion from [31] together with $\ell_1$ regularized logistic regression classification method [33] to come up with a novel angular domain channel estimator called $\ell_1$ regularized logistic regression with Toeplitz matrix reconstruction (L1-RLR-TMR) for one-bit mmWave systems. They also have employed the alternating direction method of multipliers (ADMM) [34] for solving the optimization problem of L1-RLR-TMR in an efficient manner.

Despite the significance of scalability and efficiency in one-bit massive MIMO-OFDM systems, the existing channel estimators and data detectors may not fulfill the requirement of having low computational complexity in challenging scenarios with large number of unknowns. In other words, there is a gap between the desirable computational complexity and that of the existing methods to the best of our knowledge. Therefore, the objective of this work is to fill the aforementioned gap by proposing one-bit channel estimators and data detectors that have linear order of computations with respect to the system parameters including the number of antennas at BS, the number of users, and the number of OFDM sub-carriers.

In this paper, we develop channel estimation and data detection algorithms for MIMO-OFDM systems that exploit one-bit ADCs at the BS. The channel considered here is a frequency selective channel. Inspired by outstanding properties that classification/learning-based methods have shown in solving one-bit channel estimation and data detection, we design Gaussian discriminant analysis (GDA)-based classification method [35] (known also as linear discriminant analysis (LDA)) and its approximations as so-called weak classifiers, employed in each iteration of an adaptive boosting (AdaBoost)-based scheme [33], [36]. The low computational complexity required for implementation of both GDA-based classifiers and AdaBoost make the proposed algorithms efficient, and easily scalable. In addition, flexibility in selecting the number of AdaBoost iterations enables us to gain competitive accuracy with low computational complexity.

The main contributions of our work are the following:

- An AdaBoost-based channel estimation approach for one-bit MIMO-OFDM system that operates over uncorrelated frequency selective fading channels is proposed. In each iteration of the AdaBoost-based approach, the GDA classification method along with two efficient approximations are considered as the weak classifiers. These approximate classifiers are derived by manipulating the GDA estimator. The combination of AdaBoost and GDA (and especially its approximations) enables us to estimate the channel in a remarkably efficient and yet precise manner. Specifically, using the approximations of GDA as weak classifiers at the heart of our AdaBoost approach results in having the linear order of computational complexity with respect to the problem dimension. This makes the proposed AdaBoost-based approach a versatile and also powerful tool that can be used in one-bit MIMO-OFDM systems with large number of channel entries. Numerical results validate the efficiency of the proposed AdaBoost-based channel estimator compared to other existing methods. Particularly, the AdaBoost-based channel estimator possesses similar normalized MSE (NMSE) in channel estimation as the SVM-based method of [17], whereas the computational complexity required to implement our method is substantially less than that of the SVM-based method in scenarios with large dimensions.

- We then tailor the main idea of the proposed AdaBoost-based channel estimator to fit the one-bit MIMO-OFDM data detection problem. Analogous to the proposed one-bit channel estimator, we design the data detector as an AdaBoost-based approach with considering GDA and its approximations as the weak classifiers in each iteration. The proposed one-bit data detector has desirable properties like scalability (with linear order of computations) and providing accurate data estimates. These properties are very useful in feasibility of designing one-bit MIMO-OFDM systems with high bandwidth and large number of sub-carriers. Numerical results demonstrate the strength of the proposed AdaBoost-based data detector compared to other existing methods.

The rest of the paper is organized as follows. The considered system model is presented in Section II. A brief review of GDA and AdaBoost are also presented in Section II. The proposed AdaBoost-based one-bit channel estimator and data detector are designed in Section III. Simulation results and the conclusion are presented in Section IV and Section V, respectively.

*Notation:* Upper-case and lower-case bold-face letters denote matrices and vectors, respectively, while scalars are denoted by lower-case letters. The mathematical expectation, transpose, and inverse of a square matrix are denoted by $\mathbb{E}\{\cdot\}$, $\{\cdot\}^T$, and $(\cdot)^{-1}$, respectively, while $\| \cdot \|_2$ and $\| \cdot \|_F$ denote the Euclidean norm of a vector and the Frobenius norm of a matrix. The Hadamard product is denoted by $\odot$. The $n \times n$ identity matrix is denoted by $\mathbf{I}_n$. The operator $\text{diag}\{\boldsymbol{\pi}\}$ generates a diagonal matrix by plugging the entries of the vector $\boldsymbol{\pi}$ into its main diagonal. The operators $\Re\{\cdot\}$ and $\Im\{\cdot\}$ return respectively the real and imaginary parts of the bracketed argument. The function $\mathbf{1}\{\cdot\}$ is the indicator function that is equal to 1 if its argument is true and 0 otherwise.

## II. System Model and Preliminaries

### A. One-Bit Massive MIMO-OFDM System Model

We assume a massive MIMO system comprising of $K$ users, each equipped with a single-antenna, and an $M$-antenna base station (BS) where users deploy high-resolution ADCs. Each antenna of the BS converts the real and imaginary components of the received signal from the users separately through a pair of one-bit ADCs. We specifically examine an uplink multiuser OFDM system with $N_c$ sub-carriers that operates over a frequency selective channel. The OFDM symbol in the frequency domain from the $k^{\text{th}}$ user is represented by $\mathbf{x}_k^{\text{FD}} \in \mathbb{C}^{N_c \times 1}$. To avoid confusion, we use the notations "TD" and "FD" to distinguish between time and frequency domains, respectively. We add a cyclic prefix (CP) of length $N_{\text{cp}}$ and assume that the number of channel taps $L_{\text{tap}}$ satisfies the condition $L_{\text{tap}} - 1 \le N_{\text{cp}} \le N_c$. It is assumed that $L_{\text{tap}}$ is known. Upon removing the CP, the one-bit quantized received signal at the $i^{\text{th}}$ antenna of the BS in the time domain can be expressed as follows:

$$\mathbf{y}_i^{\text{TD}} = \mathcal{Q}\left( \sum_{k=1}^{K} \mathbf{G}_{i,k}^{\text{TD}} \mathbf{F}^H \mathbf{x}_k^{\text{FD}} + \mathbf{n}_i^{\text{TD}} \right) \tag{1}$$

where $\mathbf{F} \in \mathbb{C}^{N_c \times N_c}$ denotes the normalized DFT matrix, and $\mathbf{G}_{i,k}^{\text{TD}}$ is a circulant matrix whose first column is defined by $\mathbf{g}_{i,k}^{\text{TD}} = [(\mathbf{h}_{i,k}^{\text{TD}})^T, 0, \ldots, 0]^T$. Here, $\mathbf{h}_{i,k}^{\text{TD}} \in \mathbb{C}^{L_{\text{tap}} \times 1}$ is a vector that contains the $L_{\text{tap}}$ channel taps associated with the $k^{\text{th}}$ user. The entries of $\mathbf{h}_{i,k}^{\text{TD}}$ are considered to be independent and identically distributed (i.i.d.), generated form the distribution $\mathcal{CN}\left(0, \frac{1}{L_{\text{tap}}}\right)$. Moreover, $\mathbf{n}_i^{\text{TD}} \sim \mathcal{CN}(\mathbf{0}, \mathbf{I}_{N_c})$ represents additive Gaussian noise at the $i^{\text{th}}$ antenna at the BS, whereas the notation $\mathcal{Q}(\cdot) \triangleq \text{sign}(\Re\{\cdot\}) + j\text{sign}(\Im\{\cdot\})$ represents the element-wise one-bit quantizer. The output of the operator $\text{sign}(\cdot)$ is $+1$ when the argument is a non-negative number, otherwise, the output is $-1$.

We stress here that because of the nonlinear distortion imposed by one-bit quantizers, different OFDM sub-carriers are not separable by the FFT operation as opposed to the conventional MIMO-OFDM systems. As a result, we are obliged to develop the proposed channel estimators and data detectors based on the wideband time domain representation instead of exploiting the narrowband frequency domain signals associated with each sub-carrier.

### B. Binary Classification via GDA

GDA (also known as LDA) is a classification approach that models the training examples associated with each class as samples of a normal distribution. Consider a training set that contains $m$ training examples with $n$ features and two classes denoted by $\{\mathbf{x}^{(j)}\}_{j=1,\cdots,m}$ and $y^{(j)} \in \{1, -1\}_{j=1,\cdots,m}$, respectively. GDA assumes that the corresponding training examples $\mathbf{x}^{(j)}$ for each class of $y^{(j)}$ are normally distributed with different means $\boldsymbol{\mu}_1$ and $\boldsymbol{\mu}_{-1}$, respectively, and the same covariance matrix $\boldsymbol{\Sigma}$. Therefore, depending on $y^{(j)}$, the conditional probability density function (PDF) of $\mathbf{x}^{(j)}$ can be given as one of the following equations:

$$p(\mathbf{x}^{(j)}|y^{(j)} = -1) = \frac{1}{(2\pi)^{n/2}|\boldsymbol{\Sigma}|^{1/2}} \exp\left(-\frac{1}{2}(\mathbf{x}^{(j)} - \boldsymbol{\mu}_{-1})^T \right.$$
$$\left. \times \boldsymbol{\Sigma}^{-1}(\mathbf{x}^{(j)} - \boldsymbol{\mu}_{-1})\right) \tag{2}$$

$$p(\mathbf{x}^{(j)}|y^{(j)} = 1) = \frac{1}{(2\pi)^{n/2}|\boldsymbol{\Sigma}|^{1/2}} \exp\left(-\frac{1}{2}(\mathbf{x}^{(j)} - \boldsymbol{\mu}_1)^T \right.$$
$$\left. \times \boldsymbol{\Sigma}^{-1}(\mathbf{x}^{(j)} - \boldsymbol{\mu}_1)\right) \tag{3}$$

To implement binary GDA, we need to estimate $\boldsymbol{\mu}_{-1}$, $\boldsymbol{\mu}_1$, and $\boldsymbol{\Sigma}$ from the training data. The means and the covariance matrix can be estimated as follows [35]

$$\hat{\boldsymbol{\mu}}_{-1} = \frac{\sum_{j=1}^{m} \mathbf{1}\{y^{(j)} = -1\}\mathbf{x}^{(j)}}{\sum_{j=1}^{m} \mathbf{1}\{y^{(j)} = -1\}} \tag{4}$$

$$\hat{\boldsymbol{\mu}}_1 = \frac{\sum_{j=1}^{m} \mathbf{1}\{y^{(j)} = 1\}\mathbf{x}^{(j)}}{\sum_{j=1}^{m} \mathbf{1}\{y^{(j)} = 1\}} \tag{5}$$

$$\hat{\boldsymbol{\Sigma}} = \frac{1}{m}\sum_{j=1}^{m}(\mathbf{x}^{(j)} - \hat{\boldsymbol{\mu}}_{y^{(j)}})(\mathbf{x}^{(j)} - \hat{\boldsymbol{\mu}}_{y^{(j)}})^T. \tag{6}$$

The decision boundary is then given as

$$\mathbf{h}_{\text{GDA}} = \hat{\boldsymbol{\Sigma}}^{-1}\left(\hat{\boldsymbol{\mu}}_1 - \hat{\boldsymbol{\mu}}_{-1}\right). \tag{7}$$

### C. AdaBoost

The objective of AdaBoost is to iteratively train a set of *weak classifiers* on the same data set to create a *strong classifier*. A weak classifier is identified as a classifier whose classification performance is only marginally better than random guessing. A new weak classifier is trained on a weighted version of the training data set, where the weights associated with the misclassified examples in the previous iteration are increased. Given a training set with $m$ examples, AdaBoost learns a weak classifier in the $t^{\text{th}}$ iteration which is denoted by $h_t(\mathbf{x})$. The AdaBoost algorithm is outlined in Algorithm 1. Here, $w_j^{(t)}$ is the weight of the $j^{\text{th}}$ example at the $t^{\text{th}}$ iteration, $\epsilon^{(t)}$ is the weighted error of the $t^{\text{th}}$ weak classifier, and $\alpha^{(t)}$ is the weight of the $t^{\text{th}}$ weak classifier. Moreover, $Z^{(t+1)}$ is a normalization constant that ensures that the weights sum up to 1. Despite there exists various ways to define the update rule for $w_j^{(t+1)}$, Algorithm 1 employs the exponential function to do so.

In our derivations, we use GDA and its approximate versions as weak classifiers, although there are many linear binary classifiers available in the literature that can be considered as weak classifiers. The main reason for the aforementioned choice is that these classifiers can be implemented with low computational complexities, particularly when the dimension of the unknown variables scales up.

---

**Algorithm 1** AdaBoost Algorithm

---

Input: Training set $\mathcal{S}$, number of weak classifiers $T$

Output: Final classifier $H_{\text{Ada}}$

Initialize weights $w_j^{(1)} = 1/m$ for $j = 1, 2, ..., m$

**for** $t = 1$ to $T$ **do**

    Train weak classifier $h^{(t)}(x)$ on the weighted training set $(\mathcal{S}, w)$.

    Compute error as $\epsilon^{(t)} = \sum_{j=1}^{m} w_j^{(t)} \mathbf{1}(h^{(t)}(\mathbf{x}^{(j)}) \neq y^{(j)})$.

    Compute $\alpha^{(t)} = \frac{1}{2} \ln\left(\frac{1-\epsilon^{(t)}}{\epsilon^{(t)}}\right)$.

    Update $w_j^{(t+1)} = w_j^{(t)} \exp(\alpha^{(t)} \mathbf{1}(h^{(t)}(\mathbf{x}^{(j)}) \neq y^{(j)}))$, $\forall j$.

    Compute $Z^{(t+1)} = \sum_{j=1}^{m} w_j^{(t+1)}$ and normalize weights as $w_j^{(t)} = \frac{w_j^{(t+1)}}{Z^{(t+1)}}$, $\forall j$.

**end for**

Output $H_{\text{Ada}}(\mathbf{x}) = \sum_{t=1}^{T} \alpha^{(t)} h^{(t)}(x)$

---

## III. Proposed Classification-Based Wideband Channel Estimation and Data Detection with One-bit ADCs

### A. Proposed Classification-Based Channel Estimation

For estimating the frequency selective channels explained in Section II that is utilized in the OFDM system, the frequency domain pilot vector $\mathbf{x}_k^{\text{FD}} \in \mathbb{C}^{N_c \times 1}$ is first transformed into the time domain using the inverse fast Fourier transform (IFFT) operation. The resultant time domain vector is then transmitted by the $k^{\text{th}}$ user. The one-bit quantized received signal at the $i^{\text{th}}$ antenna of the BS in (1) can be reorganized as

$$
\begin{aligned}
\mathbf{y}_i^{\text{TD}} &= \mathcal{Q}\left(\sum_{k=1}^{K} \boldsymbol{\Phi}_k^{\text{TD}} \mathbf{g}_{i,k} + \mathbf{n}_i^{\text{TD}}\right) \\
&= \mathcal{Q}\left(\sum_{k=1}^{K} \boldsymbol{\Phi}_{k,L_{\text{tap}}}^{\text{TD}} \mathbf{h}_{i,k}^{\text{TD}} + \mathbf{n}_i^{\text{TD}}\right) \\
&= \mathcal{Q}\left(\boldsymbol{\Phi}_{L_{\text{tap}}}^{\text{TD}} \mathbf{h}_i^{\text{TD}} + \mathbf{n}_i^{\text{TD}}\right)
\end{aligned}
\tag{8}
$$

where $\boldsymbol{\Phi}_k^{\text{TD}} \in \mathbb{C}^{N_c \times N_c}$ is a circulant matrix whose first column is $\boldsymbol{\phi}_k^{\text{TD}} \triangleq \mathbf{F}^H \mathbf{x}_k^{\text{FD}}$, $\boldsymbol{\Phi}_{k,L_{\text{tap}}}^{\text{TD}} \in \mathbb{C}^{N_c \times L_{\text{tap}}}$ denotes a matrix which contains only the first $L_{\text{tap}}$ columns of $\boldsymbol{\Phi}_k^{\text{TD}}$, $\boldsymbol{\Phi}_{L_{\text{tap}}}^{\text{TD}} \in \mathbb{C}^{N_c \times K L_{\text{tap}}}$ and $\mathbf{h}_i^{\text{TD}} \in \mathbb{C}^{K L_{\text{tap}} \times 1}$ respectively concatenate $\boldsymbol{\Phi}_{k,L_{\text{tap}}}^{\text{TD}}$ and $\mathbf{h}_{i,k}^{\text{TD}}$ for $k = 1, \ldots, K$ as $\boldsymbol{\Phi}_{L_{\text{tap}}}^{\text{TD}} \triangleq [\boldsymbol{\Phi}_{1,L_{\text{tap}}}^{\text{TD}}, \boldsymbol{\Phi}_{2,L_{\text{tap}}}^{\text{TD}}, \ldots, \boldsymbol{\Phi}_{K,L_{\text{tap}}}^{\text{TD}}]$ and $\mathbf{h}_i^{\text{TD}} \triangleq [(\mathbf{h}_{i,1}^{\text{TD}})^T, (\mathbf{h}_{i,2}^{\text{TD}})^T, \ldots, (\mathbf{h}_{i,K}^{\text{TD}})^T]^T$.

To simplify our derivations, we use the notation "R" as subscript when scalars, vectors, or matrices are composed of real numbers. Therefore, we transform (8) into the real domain as

$$
\mathbf{y}_{i,\text{R}}^{\text{TD}} = \text{sign}\left(\boldsymbol{\Phi}_{\text{R}}^{\text{TD}} \mathbf{h}_{i,\text{R}}^{\text{TD}} + \mathbf{n}_{i,\text{R}}^{\text{TD}}\right)
\tag{9}
$$

where

$$
\begin{aligned}
\mathbf{y}_{i,\text{R}}^{\text{TD}} &\triangleq \left[\Re\{\mathbf{y}_i^{\text{TD}}\}^T, \Im\{\mathbf{y}_i^{\text{TD}}\}^T\right]^T = \left[y_{i,\text{R},1}^{\text{TD}}, \ldots, y_{i,\text{R},2N_c}^{\text{TD}}\right]^T \\
&\in \{\pm 1\}^{2N_c \times 1}
\end{aligned}
\tag{10}
$$

$$
\begin{aligned}
\boldsymbol{\Phi}_{\text{R}} &\triangleq \begin{bmatrix} \Re\{\boldsymbol{\Phi}_{L_{\text{tap}}}^{\text{TD}}\} & -\Im\{\boldsymbol{\Phi}_{L_{\text{tap}}}^{\text{TD}}\} \\ \Im\{\boldsymbol{\Phi}_{L_{\text{tap}}}^{\text{TD}}\} & \Re\{\boldsymbol{\Phi}_{L_{\text{tap}}}^{\text{TD}}\} \end{bmatrix} \\
&= \left[\boldsymbol{\phi}_{\text{R},1}^{\text{TD}}, \boldsymbol{\phi}_{\text{R},2}^{\text{TD}}, \ldots, \boldsymbol{\phi}_{\text{R},2N_c}^{\text{TD}}\right]^T \in \mathbb{R}^{2N_c \times 2K L_{\text{tap}}}
\end{aligned}
\tag{11}
$$

$$
\mathbf{h}_{i,\text{R}}^{\text{TD}} \triangleq \left[\Re\{\mathbf{h}_i^{\text{TD}}\}^T, \Im\{\mathbf{h}_i^{\text{TD}}\}^T\right]^T \in \mathbb{R}^{2K L_{\text{tap}} \times 1}
\tag{12}
$$

$$
\mathbf{n}_{i,\text{R}}^{\text{TD}} \triangleq \left[\Re\{\mathbf{n}_i^{\text{TD}}\}^T, \Im\{\mathbf{n}_i^{\text{TD}}\}^T\right]^T \in \mathbb{R}^{2N_c \times 1}.
\tag{13}
$$

Note that $\left(\boldsymbol{\phi}_{\text{R},j}^{\text{TD}}\right)^T$ with $j \in \{1, 2, \ldots, 2N_c\}$ is the $j^{\text{th}}$ row of $\boldsymbol{\Phi}_{\text{R}}$ here. Additionally, as suggested by (12), estimating $\{\mathbf{h}_i^{\text{TD}}\}_{i=1,2,\ldots,M}$ is equivalent to estimating $\{\mathbf{h}_{i,\text{R}}^{\text{TD}}\}_{i=1,2,\ldots,M}$.

We emphasize that binary classification methods can be employed for estimating $\mathbf{h}_{i,\text{R}}^{\text{TD}}$ in (9). Here, $\boldsymbol{\phi}_{\text{R},j}^{\text{TD}}$ and $y_{i,\text{R},j}^{\text{TD}}$ with $j \in \{1, 2, \ldots, 2N_c\}$ serve as the training examples and the class labels, respectively. In other words, (9)-(12) can be viewed as a binary classification problem with the training set $\mathcal{S}_i = \{\mathbf{x}^{(j)} = \boldsymbol{\phi}_{\text{R},j}^{\text{TD}}, y^{(j)} = y_{i,\text{R},j}^{\text{TD}}\}_{j=1,2,\ldots,2N_c}$ and the decision boundary $\mathbf{h}_{i,\text{R}}^{\text{TD}}$ based on the definitions provided in the prequel. Hence, we can exploit the GDA classification method as the weak classifier in each iteration of an AdaBoost-based approach for estimating $\mathbf{h}_{i,\text{R}}^{\text{TD}}$. The computation of the means and covariance matrix (4)-(6) then should be revised in the $t^{\text{th}}$ iteration of the proposed AdaBoost-based approach as

$$
\hat{\boldsymbol{\mu}}_{-1}^{(t)} = \sum_{j=1}^{2N_c} \mathbf{1}\{y_{i,\text{R},j}^{\text{TD}} = -1\} w_j^{(t)} \boldsymbol{\phi}_{\text{R},j}^{\text{TD}}
\tag{14}
$$

$$
\hat{\boldsymbol{\mu}}_1^{(t)} = \sum_{j=1}^{2N_c} \mathbf{1}\{y_{i,\text{R},j}^{\text{TD}} = 1\} w_j^{(t)} \boldsymbol{\phi}_{\text{R},j}^{\text{TD}}
\tag{15}
$$

$$
\hat{\boldsymbol{\Sigma}}^{(t)} = \sum_{j=1}^{2N_c} w_j^{(t)} (\boldsymbol{\phi}_{\text{R},j}^{\text{TD}} - \hat{\boldsymbol{\mu}}_{y_{i,\text{R},j}^{\text{TD}}}^{(t)}) (\boldsymbol{\phi}_{\text{R},j}^{\text{TD}} - \hat{\boldsymbol{\mu}}_{y_{i,\text{R},j}^{\text{TD}}}^{(t)})^T
\tag{16}
$$

$$
\hat{\mathbf{h}}_{i,\text{R}}^{\text{TD},(t)} = \left(\hat{\boldsymbol{\Sigma}}^{(t)}\right)^{-1} \left(\hat{\boldsymbol{\mu}}_1^{(t)} - \hat{\boldsymbol{\mu}}_{-1}^{(t)}\right)
\tag{17}
$$

where $w_j^{(t)}$ represents the weight of the $j^{\text{th}}$ training example at the $t^{\text{th}}$ iteration.

To implement (17), the inverse of the matrix $\hat{\boldsymbol{\Sigma}}^{(t)}$ should be calculated, which requires the computational complexity of $\mathcal{O}\left((K L_{\text{tap}})^{2.373}\right)$. This computational complexity can considerably restrict the time efficiency of implementing (17), especially when the multiplication of $K$ and $L_{\text{tap}}$ grows larger. At the same time, as a weak classifier is required to be slightly better than random guesses, the accurate knowledge of the inverse of $\hat{\boldsymbol{\Sigma}}^{(t)}$ is not needed. Thus, it is reasonable to consider approximating (17) to avoid the cubic computational complexity of calculating $\left(\hat{\boldsymbol{\Sigma}}^{(t)}\right)^{-1}$. Towards this end, two approximations of (17) are introduced in the following.

*Approximation 1*: As the first approximation, we propose to modify (16) as

$$
\hat{\boldsymbol{\Sigma}}_1^{(t)} \triangleq \text{diag}\left\{\hat{\boldsymbol{\sigma}}_1^{(t)}\right\}
\tag{18}
$$

where

$$\hat{\boldsymbol{\sigma}}_1^{(t)} = \sum_{j=1}^{2N_c} w_j^{(t)} \left( (\boldsymbol{\phi}_{R,j}^{TD} - \hat{\boldsymbol{\mu}}_{y_{i,R,j}^{(t)}}^{TD}) \odot (\boldsymbol{\phi}_{R,j}^{TD} - \hat{\boldsymbol{\mu}}_{y_{i,R,j}^{(t)}}^{TD}) \right) \quad (19)$$

The essence of this approximation is to set all off-diagonal elements of $\hat{\boldsymbol{\Sigma}}^{(t)}$ in (16) to zero and preserve only its diagonal elements. In other words, only the diagonal elements of the original matrix $\hat{\boldsymbol{\Sigma}}^{(t)}$ in (16) need to be computed as the vector $\hat{\boldsymbol{\sigma}}_1^{(t)}$ in (19), and $\hat{\boldsymbol{\Sigma}}_1^{(t)}$ is defined using $\hat{\boldsymbol{\sigma}}_1^{(t)}$ as in (18). Then, (17) is modified as

$$\hat{\mathbf{h}}_{i,R,app1}^{TD,(t)} \triangleq \left( \hat{\boldsymbol{\Sigma}}_1^{(t)} \right)^{-1} \left( \hat{\boldsymbol{\mu}}_1^{(t)} - \hat{\boldsymbol{\mu}}_{-1}^{(t)} \right) \quad (20)$$

Note that the use of $\hat{\boldsymbol{\Sigma}}_1^{(t)}$ instead of the original $\hat{\boldsymbol{\Sigma}}^{(t)}$ considerably reduces the computational complexity of computing $\hat{\mathbf{h}}_{i,R}^{TD,(t)}$.

*Approximation 2:* We propose to set $\hat{\boldsymbol{\Sigma}}^{(t)} = \mathbf{I}_{2KL_{tap}}$ in (17). Then, the modified estimate of $\mathbf{h}_{i,R}^{TD,(t)}$ is expressed as

$$\hat{\mathbf{h}}_{i,R,app2}^{TD,(t)} \triangleq \hat{\boldsymbol{\mu}}_1^{(t)} - \hat{\boldsymbol{\mu}}_{-1}^{(t)} \quad (21)$$

where the weak classifier of (17) is approximated as the distance between the mean vectors of the two classes in (21). We stress here that the later requires substantially less computations compared to that of the former.

The steps of the proposed methods are outlined in Algorithm 2.[1] It should be noted that Algorithm 1 presents the generic procedure of the AdaBoost approach for using weak binary classifiers/learners $h^{(t)}(x)$ to build a strong binary classifier/learner $H_{Ada}(x)$, whereas we exploit the core idea of AdaBoost to use the weak channel estimates $\mathbf{h}_i^{(t)}$ to build a strong channel estimate $\hat{\mathbf{h}}_{i,R}^{TD}$ in Algorithm 2. We emphasize here the difference of $h^{(t)}(x)$ and $H_{Ada}(x)$ with $\mathbf{h}_i^{(t)}$ and $\hat{\mathbf{h}}_{i,R}^{TD}$, that is, the former represents binary classifier while the later denotes the separating hyperplane in a binary classification problem.

Note that a normalization step is applied to the output of the AdaBoost-based methods outlined in Algorithm 2. The reason for this is that the estimates provided by these methods only specify the direction of $\mathbf{h}_{i,R}^{TD}$, while the magnitude remains unknown since the one-bit ADCs preserve only the sign of the received signals. Therefore, $\beta \mathbf{h}_{i,R}^{TD}$ for any $\beta > 0$ will yield the same $\mathbf{y}_{i,R}^{TD}$ as in (10). Here, since we assume that $2KL_{tap}$ elements of $\mathbf{h}_{i,R}^{TD}$ are independent with variance $1/(2L_{tap})$, the last normalization step is added to ensure that the channel estimates have squared norm of $K$.

**Remark 1:** To ensure the clarity of presentation, we used a loop to estimate $\mathbf{h}_{i,R}^{TD}$ for $i \in \{1, 2, \dots, M\}$ in Algorithm 2. However, it is important to note that these $M$ channel vectors can be estimated in parallel, resulting in a reduction in the overall run time of the channel estimation procedure.

[1] Although $\alpha^{(t)} = \frac{1}{2} \ln \left( \frac{1-\epsilon^{(t)}}{\epsilon^{(t)}} \right)$ in the original AdaBoost algorithm (refer to Algorithm 1), we have observed that setting $\alpha^{(t)} = \frac{1}{4} \ln \left( \frac{1-\epsilon^{(t)}}{\epsilon^{(t)}} \right)$ in Algorithms 2 and 3 results in better performance for the problems solved in this paper.

---

**Algorithm 2** One-bit GDA-AdaBoost Algorithms for Channel Estimation

**Input:** $\mathcal{S}_i = \{\mathbf{x}^{(j)} = \boldsymbol{\phi}_{R,j}^{TD}, y^{(j)} = y_{i,R,j}^{TD}\}_{j=1,2,\dots,2N_c}$ for $i \in \{1, 2, \dots, M\}$ whose elements are defined in (10) and (11), and number of weak classifiers $T$.

**Output:** $\hat{\mathbf{h}}_{i,R}^{TD}$ for $i \in \{1, 2, \dots, M\}$.

**for** $i = 1$ **to** $M$ **do**

  Initialize weights $w_j^{(1)} = \frac{1}{2N_c}$ for $j \in \{1, 2, \dots, 2N_c\}$.

  **for** $t = 1$ **to** $T$ **do**

    Use the training set $\mathcal{S}_i$ to compute $\hat{\boldsymbol{\mu}}_{-1}^{(t)}$, $\hat{\boldsymbol{\mu}}_1^{(t)}$, $\hat{\boldsymbol{\Sigma}}^{(t)}$, and $\hat{\boldsymbol{\Sigma}}_1^{(t)}$ via (14)-(16) and (18), respectively. Then, compute the $t^{th}$ weak classifier as:

    **one-bit GDA-Ada**
$$\mathbf{h}_i^{(t)} = \left( \hat{\boldsymbol{\Sigma}}^{(t)} \right)^{-1} \left( \hat{\boldsymbol{\mu}}_1^{(t)} - \hat{\boldsymbol{\mu}}_{-1}^{(t)} \right)$$

    **one-bit GDA-Ada-1**
$$\mathbf{h}_i^{(t)} = \left( \hat{\boldsymbol{\Sigma}}_1^{(t)} \right)^{-1} \left( \hat{\boldsymbol{\mu}}_1^{(t)} - \hat{\boldsymbol{\mu}}_{-1}^{(t)} \right)$$

    **one-bit GDA-Ada-2**
$$\mathbf{h}_i^{(t)} = \hat{\boldsymbol{\mu}}_1^{(t)} - \hat{\boldsymbol{\mu}}_{-1}^{(t)}.$$

    Compute error as
$$\epsilon^{(t)} = \sum_{j=1}^{2N_c} w_j^{(t)} \mathbf{1} \left( (\boldsymbol{\phi}_{R,j}^{TD})^T \mathbf{h}_i^{(t)} \neq y^{(j)} \right).$$

    Compute $\alpha^{(t)} = \frac{1}{4} \ln \left( \frac{1-\epsilon^{(t)}}{\epsilon^{(t)}} \right)$.

    Update $w_j^{(t+1)} = w_j^{(t)} \exp \left( \alpha^{(t)} \mathbf{1} \left( (\boldsymbol{\phi}_{R,j}^{TD})^T \mathbf{h}_i^{(t)} \neq y^{(j)} \right) \right)$, $\forall j$.

    Compute $Z^{(t+1)} = \sum_{j=1}^{2N_c} w_j^{(t+1)}$ and normalize weights as $w_j^{(t+1)} = \frac{w_j^{(t+1)}}{Z^{(t+1)}}$, $\forall j$.

  **end for**

  Construct $\tilde{\mathbf{h}}_{i,R}^{TD} = \sum_{t=1}^T \alpha^{(t)} \mathbf{h}_i^{(t)}$, and then normalize as $\hat{\mathbf{h}}_{i,R}^{TD} = \frac{\sqrt{K} \tilde{\mathbf{h}}_{i,R}^{TD}}{\|\tilde{\mathbf{h}}_{i,R}^{TD}\|_2}$.

**end for**

---

**Remark 2:** The key feature of AdaBoost that allows us to approximate (17) as (20) and (21) without sacrificing estimation performance is that it can incorporate weak classifiers that are only slightly better than random guessing and combine them to form a strong classifier. The approximation of (20) and (21) are justifiable because they are certainly better than random guessing, hence they can be treated as weak classifiers. In this regard, AdaBoost is a powerful approach to build a strong classifier out of weak classifiers with low computational complexity.

### B. Proposed Classification-Based Data Detection

In this section, we propose AdaBoost-based methods for one-bit data detection in OFDM systems with frequency selective channels. To begin with, the one-bit quantized received signal at the $i^{th}$ antenna of the BS in (1) can be rewritten as

$$\mathbf{y}_i^{TD} = \mathcal{Q} \left( \sum_{k=1}^K \mathbf{G}_{i,k}^{TD} \mathbf{F}^H \mathbf{x}_k^{FD} + \mathbf{n}_i^{TD} \right)$$
$$= \mathcal{Q} \left( \mathbf{G}_i^{FD} \mathbf{x}^{FD} + \mathbf{n}_i^{TD} \right) \quad (22)$$

where $\mathbf{G}_i^{FD} \triangleq [\mathbf{G}_{i,1}^{TD} \mathbf{F}^H, \dots, \mathbf{G}_{i,K}^{TD} \mathbf{F}^H] \in \mathbb{C}^{N_c \times KN_c}$ and $\mathbf{x}^{FD} \triangleq [(\mathbf{x}_1^{FD})^T, (\mathbf{x}_2^{FD})^T, \dots, (\mathbf{x}_K^{FD})^T]^T \in \mathbb{C}^{KN_c \times 1}$. The

former represents the pre-estimated/known CSI, while the later is the symbol vectors transmitted over $N_c$ subcarriers by the $K$ users. The objective here is to recover the vector $\mathbf{x}^{\mathrm{FD}}$ and then identify the symbols transmitted. Placing all $\{\mathbf{y}_i^{\mathrm{TD}}\}_{i=1,2,\ldots,M}$ in a vector as $\mathbf{y}^{\mathrm{TD}} \triangleq [(\mathbf{y}_1^{\mathrm{TD}})^T, (\mathbf{y}_2^{\mathrm{TD}})^T, \ldots, (\mathbf{y}_M^{\mathrm{TD}})^T]^T \in \mathbb{C}^{MN_c \times 1}$, we obtain

$$\mathbf{y}^{\mathrm{TD}} = \mathcal{Q}\left(\mathbf{G}^{\mathrm{FD}}\mathbf{x}^{\mathrm{FD}} + \mathbf{n}^{\mathrm{TD}}\right) \tag{23}$$

where $\mathbf{G}^{\mathrm{FD}} \triangleq [(\mathbf{G}_1^{\mathrm{FD}})^T, (\mathbf{G}_2^{\mathrm{FD}})^T, \ldots, (\mathbf{G}_M^{\mathrm{FD}})^T]^T \in \mathbb{C}^{MN_c \times KN_c}$. The real domain transformation of (23) is given as

$$\mathbf{y}_{\mathrm{R}}^{\mathrm{TD}} = \mathrm{sign}\left(\mathbf{G}_{\mathrm{R}}^{\mathrm{FD}}\mathbf{x}_{\mathrm{R}}^{\mathrm{FD}} + \mathbf{n}_{\mathrm{R}}^{\mathrm{TD}}\right) \tag{24}$$

where

$$\mathbf{y}_{\mathrm{R}}^{\mathrm{TD}} \triangleq \left[\Re\{\mathbf{y}^{\mathrm{TD}}\}^T, \Im\{\mathbf{y}^{\mathrm{TD}}\}^T\right]^T = \left[y_{\mathrm{R},1}^{\mathrm{TD}}, \ldots, y_{\mathrm{R},2MN_c}^{\mathrm{TD}}\right]^T$$
$$\in \{\pm 1\}^{2MN_c \times 1} \tag{25}$$

$$\mathbf{G}_{\mathrm{R}}^{\mathrm{FD}} \triangleq \begin{bmatrix} \Re\{\mathbf{G}^{\mathrm{FD}}\} & -\Im\{\mathbf{G}^{\mathrm{FD}}\} \\ \Im\{\mathbf{G}^{\mathrm{FD}}\} & \Re\{\mathbf{G}^{\mathrm{FD}}\} \end{bmatrix}$$
$$= \left[\mathbf{g}_{\mathrm{R},1}^{\mathrm{FD}}, \mathbf{g}_{\mathrm{R},2}^{\mathrm{FD}}, \ldots, \mathbf{g}_{\mathrm{R},2MN_c}^{\mathrm{FD}}\right]^T \in \mathbb{R}^{2MN_c \times 2KN_c} \tag{26}$$

$$\mathbf{x}_{\mathrm{R}}^{\mathrm{FD}} \triangleq \left[\Re\{\mathbf{x}^{\mathrm{FD}}\}^T, \Im\{\mathbf{x}^{\mathrm{FD}}\}^T\right]^T \in \mathbb{R}^{2KN_c \times 1} \tag{27}$$

$$\mathbf{n}_{\mathrm{R}}^{\mathrm{TD}} \triangleq \left[\Re\{\mathbf{n}^{\mathrm{TD}}\}^T, \Im\{\mathbf{n}^{\mathrm{TD}}\}^T\right]^T \in \mathbb{R}^{2MN_c \times 1}. \tag{28}$$

Here $\left\{\left(\mathbf{g}_{\mathrm{R},j}^{\mathrm{FD}}\right)^T\right\}_{j=1,\ldots,2MN_c}$ is the $j^{\mathrm{t}}$ row of $\mathbf{G}_{\mathrm{R}}^{\mathrm{FD}}$.

Analogous to the problem of estimating $\mathbf{h}_{i,\mathrm{R}}^{\mathrm{TD}}$ in (9), the problem of estimating $\mathbf{x}_{\mathrm{R}}^{\mathrm{FD}}$ in (24) can be treated as a binary classification problem where $\mathbf{x}_{\mathrm{R}}^{\mathrm{FD}}$ serves as the separating hyperplane between two classes. Therefore, we can constitute the binary classification training set as $\mathcal{S}_{\mathrm{d}} = \{\mathbf{x}^{(j)} = \mathbf{g}_{\mathrm{R},j}^{\mathrm{FD}}, y^{(j)} = y_{\mathrm{R},j}^{\mathrm{TD}}\}_{j=1,2,\ldots,2MN_c}$ based on (24)-(26) with the aim of estimating $\mathbf{x}_{\mathrm{R}}^{\mathrm{FD}}$ as the corresponding separating hyperplane. Thus, the GDA classification method along with two approximations derived in Subsection III-A can be used as weak classifiers in each iteration of an AdaBoost-based approach for recovering $\mathbf{x}_{\mathrm{R}}^{\mathrm{FD}}$. In this regard, the counterparts of (17), (20), and (21) with respect to $\mathbf{x}_{\mathrm{R}}^{\mathrm{FD}}$ are respectively expressed as

$$\hat{\mathbf{x}}_{\mathrm{d}}^{\mathrm{FD},(t)} = \left(\hat{\boldsymbol{\Sigma}}_{\mathrm{d}}^{(t)}\right)^{-1}\left(\hat{\boldsymbol{\mu}}_{\mathrm{d},1}^{(t)} - \hat{\boldsymbol{\mu}}_{\mathrm{d},-1}^{(t)}\right) \tag{29}$$

$$\hat{\mathbf{x}}_{\mathrm{d,app1}}^{(t)} = \left(\hat{\boldsymbol{\Sigma}}_{\mathrm{d},1}^{(t)}\right)^{-1}\left(\hat{\boldsymbol{\mu}}_{\mathrm{d},1}^{(t)} - \hat{\boldsymbol{\mu}}_{\mathrm{d},-1}^{(t)}\right) \tag{30}$$

$$\hat{\mathbf{x}}_{\mathrm{d,app2}}^{(t)} = \hat{\boldsymbol{\mu}}_{\mathrm{d},1}^{(t)} - \hat{\boldsymbol{\mu}}_{\mathrm{d},-1}^{(t)} \tag{31}$$

where

$$\hat{\boldsymbol{\mu}}_{\mathrm{d},-1}^{(t)} = \sum_{j=1}^{2MN_c} \mathbf{1}\{y_{\mathrm{R},j}^{\mathrm{TD}} = -1\} w_j^{(t)} \mathbf{g}_{\mathrm{R},j}^{\mathrm{FD}} \tag{32}$$

$$\hat{\boldsymbol{\mu}}_{\mathrm{d},1}^{(t)} = \sum_{j=1}^{2MN_c} \mathbf{1}\{y_{\mathrm{R},j}^{\mathrm{TD}} = 1\} w_j^{(t)} \mathbf{g}_{\mathrm{R},j}^{\mathrm{FD}} \tag{33}$$

$$\hat{\boldsymbol{\Sigma}}_{\mathrm{d}}^{(t)} = \sum_{j=1}^{2MN_c} w_j^{(t)} (\mathbf{g}_{\mathrm{R},j}^{\mathrm{FD}} - \hat{\boldsymbol{\mu}}_{\mathrm{d},y_{\mathrm{R},j}^{\mathrm{TD}}}^{(t)})(\mathbf{g}_{\mathrm{R},j}^{\mathrm{FD}} - \hat{\boldsymbol{\mu}}_{\mathrm{d},y_{\mathrm{R},j}^{\mathrm{TD}}}^{(t)})^T \tag{34}$$

$$\hat{\boldsymbol{\Sigma}}_{\mathrm{d},1}^{(t)} = \mathrm{diag}\left\{\hat{\boldsymbol{\sigma}}_{\mathrm{d},1}^{(t)}\right\} \tag{35}$$

$$\hat{\boldsymbol{\sigma}}_{\mathrm{d},1}^{(t)} = \sum_{j=1}^{2MN_c} w_j^{(t)}\left((\mathbf{g}_{\mathrm{R},j}^{\mathrm{FD}} - \hat{\boldsymbol{\mu}}_{\mathrm{d},y_{\mathrm{R},j}^{\mathrm{TD}}}^{(t)}) \odot (\mathbf{g}_{\mathrm{R},j}^{\mathrm{FD}} - \hat{\boldsymbol{\mu}}_{\mathrm{d},y_{\mathrm{R},j}^{\mathrm{TD}}}^{(t)})\right). \tag{36}$$

Note that $w_j^{(t)}$ is the weight assigned to the $j^{\mathrm{th}}$ training example in the $t^{\mathrm{th}}$ iteration. In addition, the notation "d" is used as subscript in (29)-(36) to avoid confusion with channel estimation part's of equations. Let $\mathbf{x}_{\mathrm{d}}^{(t)}$ represent the estimated signal in the $t^{\mathrm{th}}$ iteration using either one of the weak classifiers in (29)-(31). A normalization step is needed to match the power of the estimated signal with that of the actual transmitted signal.[2] Then, we have

$$\bar{\mathbf{x}}_{\mathrm{d}}^{(t)} = \frac{\sqrt{KN_c}\mathbf{x}_{\mathrm{d}}^{(t)}}{\|\mathbf{x}_{\mathrm{d}}^{(t)}\|_2} = [\bar{x}_{\mathrm{d},1}^{(t)}, \bar{x}_{\mathrm{d},2}^{(t)}, \ldots, \bar{x}_{\mathrm{d},2KN_c}^{(t)}]^T. \tag{37}$$

The next step is to map/project the elements of $\bar{\mathbf{x}}_{\mathrm{d}}^{(t)}$ to one member of the transmitted signal constellations set denoted by $\mathcal{F}$ by solving the following optimization problem symbol-by-symbol:

$$\tilde{x}_{\mathrm{d},k}^{(t)} = \arg\min_{x \in \mathcal{F}} |x - (\bar{x}_{\mathrm{d},k}^{(t)} + j\bar{x}_{\mathrm{d},k+KN_c}^{(t)})|$$
$$\text{for } k = 1, 2, \ldots, KN_c \tag{38}$$

where $\tilde{x}_{\mathrm{d},k}^{(t)}$ is the $k^{\mathrm{th}}$ entry of the estimated signal in the $t^{\mathrm{th}}$ iteration. Thus, the signal vector is $\tilde{\mathbf{x}}_{\mathrm{d}}^{(t)} = [\tilde{x}_{\mathrm{d},1}^{(t)}, \tilde{x}_{\mathrm{d},2}^{(t)}, \ldots, \tilde{x}_{\mathrm{d},KN_c}^{(t)}]^T$.

Transforming $\tilde{\mathbf{x}}_{\mathrm{d}}^{(t)}$ into the real domain as $\tilde{\mathbf{x}}_{\mathrm{d}}^{(t)} = [\Re\{\tilde{\mathbf{x}}_{\mathrm{d}}^{(t)}\}^T, \Im\{\tilde{\mathbf{x}}_{\mathrm{d}}^{(t)}\}^T]^T$, we can obtain $\epsilon^{(t)}$, $\alpha^{(t)}$, and $w_j^{(t+1)}$ for $j = 1, 2, \ldots, 2MN_c$. After executing $T$ iterations, the AdaBoost outputs $\tilde{\mathbf{x}}_{\mathrm{R}}^{\mathrm{FD}} = \sum_{t=1}^{T} \alpha^{(t)}\tilde{\mathbf{x}}_{\mathrm{d}}^{(t)}$. Analogous to (37) and (38), the final steps are to first normalize $\tilde{\mathbf{x}}_{\mathrm{R}}^{\mathrm{FD}}$, and then perform the symbol-by-symbol mapping as follows

$$\bar{\mathbf{x}}_{\mathrm{R}}^{\mathrm{FD}} = \frac{\sqrt{KN_c}\tilde{\mathbf{x}}_{\mathrm{R}}^{\mathrm{FD}}}{\|\tilde{\mathbf{x}}_{\mathrm{R}}^{\mathrm{FD}}\|_2} = [\bar{x}_{\mathrm{R},1}^{\mathrm{FD}}, \bar{x}_{\mathrm{R},2}^{\mathrm{FD}}, \ldots, \bar{x}_{\mathrm{R},2KN_c}^{\mathrm{FD}}]^T \tag{39}$$

$$\hat{x}_k^{\mathrm{FD}} = \arg\min_{x \in \mathcal{F}} |x - (\bar{x}_{\mathrm{R},k}^{\mathrm{FD}} + j\bar{x}_{\mathrm{R},k+KN_c}^{\mathrm{FD}})|$$
$$\text{for } k = 1, 2, \ldots, KN_c \tag{40}$$

where $\hat{x}_k^{\mathrm{FD}}$ is the $k^{\mathrm{th}}$ entry of the final estimate. Thus, the final estimate is $\hat{\mathbf{x}}^{\mathrm{FD}} \triangleq [\hat{x}_1^{\mathrm{FD}}, \hat{x}_2^{\mathrm{FD}}, \ldots, \hat{x}_{KN_c}^{\mathrm{FD}}]^T$. The steps of the proposed data detection methods are listed in Algorithm 3.

We emphasize here that the loops associated with (38) and (40) are only included for the sake of presentation clarity in Algorithm 3, and the symbol-by-symbol detection can be executed concurrently. It is also worth noting that post-processing can be performed for refining the outputs of (40) as have been suggested in [8] and [17]. The former has exploited the ML criterion to select the final data symbol from a properly designed data candidate set [8], whereas the latter has resorted to a *minimum weighted Hamming distance*-based criterion [37] to pick up the refined data symbol from a data candidate set. Despite the efficiency of the aforementioned post-processing data refinement, we do not use it here and the simulation results for the proposed methods are provided without considering the post-processing in the next section.

**Remark 3:** One of the advantages of the proposed AdaBoost-based algorithms is that the sufficient number of weak classifiers for obtaining a reasonable accuracy is of order

---

[2]Such normalization is also used in [8] and [17] for example.

**Algorithm 3** One-bit GDA-AdaBoost Algorithms for data detection

---

**Input:** $\mathcal{S}_{\mathrm{d}} = \{\mathbf{x}^{(j)} = \mathbf{g}_{\mathrm{R},j}^{\mathrm{FD}}, y^{(j)} = y_{\mathrm{R},j}^{\mathrm{TD}}\}_{j=1,2,\ldots,2MN_{\mathrm{c}}}$ whose elements are defined in (25) and (26), and number of weak classifiers $T$.

**Output:** $\hat{\mathbf{x}}^{\mathrm{FD}}$.

Initialize $w_j^{(1)} = \frac{1}{2MN_{\mathrm{c}}}$ for $j \in \{1, 2, \ldots, 2MN_{\mathrm{c}}\}$.

**for** $t = 1$ to $T$ **do**

   Use the training set $\mathcal{S}_{\mathrm{d}}$ to compute $\hat{\boldsymbol{\mu}}_{\mathrm{d},-1}^{(t)}$, $\hat{\boldsymbol{\mu}}_{\mathrm{d},1}^{(t)}$, $\hat{\boldsymbol{\Sigma}}_{\mathrm{d}}^{(t)}$, and $\hat{\boldsymbol{\Sigma}}_{\mathrm{d},1}^{(t)}$ via (32)-(35), respectively. Then, compute the $t^{\mathrm{th}}$ weak classifier as:

   **one-bit GDA-Ada**
   $$\mathbf{x}_{\mathrm{d}}^{(t)} = \left(\hat{\boldsymbol{\Sigma}}_{\mathrm{d}}^{(t)}\right)^{-1}\left(\hat{\boldsymbol{\mu}}_{\mathrm{d},1}^{(t)} - \hat{\boldsymbol{\mu}}_{\mathrm{d},-1}^{(t)}\right)$$
   **one-bit GDA-Ada-1**
   $$\mathbf{x}_{\mathrm{d}}^{(t)} = \left(\hat{\boldsymbol{\Sigma}}_{\mathrm{d},1}^{(t)}\right)^{-1}\left(\hat{\boldsymbol{\mu}}_{\mathrm{d},1}^{(t)} - \hat{\boldsymbol{\mu}}_{\mathrm{d},-1}^{(t)}\right)$$
   **one-bit GDA-Ada-2**
   $$\mathbf{x}_{\mathrm{d}}^{(t)} = \hat{\boldsymbol{\mu}}_{\mathrm{d},1}^{(t)} - \hat{\boldsymbol{\mu}}_{\mathrm{d},-1}^{(t)}.$$
   Normalize $\mathbf{x}_{\mathrm{d}}^{(t)}$ as $\bar{\mathbf{x}}_{\mathrm{d}}^{(t)} = \frac{\sqrt{KN_{\mathrm{c}}}\mathbf{x}_{\mathrm{d}}^{(t)}}{\|\mathbf{x}_{\mathrm{d}}^{(t)}\|_2}$, and denote the $k^{\mathrm{th}}$ entry of $\bar{\mathbf{x}}_{\mathrm{d}}^{(t)}$ as $\bar{x}_{\mathrm{d},k}^{(t)}$ for $k \in \{1, 2, \ldots, 2KN_{\mathrm{c}}\}$.

   **for** $k' = 1$ to $KN_{\mathrm{c}}$ **do**
      Solve the optimization problem (38) to detect $\tilde{x}_{\mathrm{d},k'}^{(t)}$.
   **end for**

   Construct $\tilde{\mathbf{x}}_{\mathrm{d}}^{(t)} = [\tilde{x}_{\mathrm{d},1}^{(t)}, \tilde{x}_{\mathrm{d},2}^{(t)}, \ldots, \tilde{x}_{\mathrm{d},KN_{\mathrm{c}}}^{(t)}]^T$ and $\check{\mathbf{x}}_{\mathrm{d}}^{(t)} = [\Re\{\tilde{\mathbf{x}}_{\mathrm{d}}^{(t)}\}^T, \Im\{\tilde{\mathbf{x}}_{\mathrm{d}}^{(t)}\}^T]^T$.

   Compute error as
   $$\epsilon^{(t)} = \sum_{j=1}^{2MN_{\mathrm{c}}} w_j^{(t)} \mathbf{1}\left((\mathbf{g}_{\mathrm{R},j}^{\mathrm{FD}})^T\check{\mathbf{x}}_{\mathrm{d}}^{(t)} \neq y^{(j)}\right).$$
   Compute $\alpha^{(t)} = \frac{1}{4}\ln\left(\frac{1-\epsilon^{(t)}}{\epsilon^{(t)}}\right)$.
   Update $w_j^{(t+1)} = w_j^{(t)}\exp\left(\alpha^{(t)}\mathbf{1}\left((\mathbf{g}_{\mathrm{R},j}^{\mathrm{FD}})^T\check{\mathbf{x}}_{\mathrm{d}}^{(t)} \neq y^{(j)}\right)\right)$, $\forall j$.
   Compute $Z^{(t+1)} = \sum_{j=1}^{2MN_{\mathrm{c}}} w_j^{(t+1)}$ and normalize weights as $w_j^{(t+1)} = \frac{w_j^{(t+1)}}{Z^{(t+1)}}$, $\forall j$.

**end for**

Construct $\check{\mathbf{x}}_{\mathrm{R}}^{\mathrm{FD}} = \sum_{t=1}^{T} \alpha^{(t)}\check{\mathbf{x}}_{\mathrm{d}}^{(t)}$, and then normalize as $\bar{\mathbf{x}}_{\mathrm{R}}^{\mathrm{FD}} = \frac{\sqrt{KN_{\mathrm{c}}}\check{\mathbf{x}}_{\mathrm{R}}^{\mathrm{FD}}}{\|\check{\mathbf{x}}_{\mathrm{R}}^{\mathrm{FD}}\|_2}$. Denote the $k^{\mathrm{th}}$ entry of $\bar{\mathbf{x}}_{\mathrm{R}}^{\mathrm{FD}}$ as $\bar{x}_{\mathrm{R},k}^{\mathrm{FD}}$ for $k \in \{1, 2, \ldots, 2KN_{\mathrm{c}}\}$.

**for** $k' = 1$ to $KN_{\mathrm{c}}$ **do**
   Solve the optimization problem (40) to detect $\hat{x}_{k'}^{\mathrm{FD}}$.
**end for**

Construct $\hat{\mathbf{x}}^{\mathrm{FD}} = [\hat{x}_1^{\mathrm{FD}}, \hat{x}_2^{\mathrm{FD}}, \ldots, \hat{x}_{KN_{\mathrm{c}}}^{\mathrm{FD}}]^T$.

---

of a few tens. In other words, increasing $T$ over just a few tens does not change the performance of the proposed AdaBoost-based algorithms dramatically. We recommend to set $T = 10$ for the proposed channel estimator and data detector because this value has been found to be effective in reaching accurate results. The impact of using different values of $T$ for channel estimation and data detection will be examined in the next section though.

### C. Computational Complexity

Implementing one-bit GDA-Ada, one-bit GDA-Ada-1, and one-bit GDA-Ada-2 channel estimators described in Algorithm 2 require $\mathcal{O}\left(TM\max\{(KL_{\mathrm{tab}})^{2.373}, (KL_{\mathrm{tab}})^2N_{\mathrm{c}}\}\right)$,

$\mathcal{O}\left(TMKL_{\mathrm{tab}}N_{\mathrm{c}}\right)$, and $\mathcal{O}\left(TMKL_{\mathrm{tab}}N_{\mathrm{c}}\right)$ flops, respectively. Noteworthy to mention that one-bit GDA-Ada-1 and one-bit GDA-Ada-2 channel estimators have first-order (linear) theoretical computational complexity with respect to $M$, $K$, $L_{\mathrm{tab}}$, and $N_{\mathrm{c}}$, which is analogous to Bayesian-based methods [13], [14] (see Table I). However, it will be shown in the next section that the run times required for implementing one-bit GDA-Ada-1 and one-bit GDA-Ada-2 channel estimators are significantly lower than the run time required for implementing the BiGAMP-based channel estimator though (see Fig. 2). Moreover, the computations for one-bit GDA-Ada-1 and one-bit GDA-Ada-2 channel estimators can be straightforwardly parallelized.

In addition, implementing one-bit GDA-Ada, one-bit GDA-Ada-1, and one-bit GDA-Ada-2 data detectors presented in Algorithm 3 require $\mathcal{O}\left(T\max\{(KN_{\mathrm{c}})^{2.373}, MK^2N_{\mathrm{c}}^3\}\right)$, $\mathcal{O}\left(TMKN_{\mathrm{c}}^2\right)$, and $\mathcal{O}\left(TMKN_{\mathrm{c}}^2\right)$, respectively. Similar to the channel estimation case, one-bit GDA-Ada-1 and one-bit GDA-Ada-2 data detectors have first-order (linear) computational complexity with respect to $M$ and $K$. However, they have second-order (quadratic) computational complexity with respect to $N_{\mathrm{c}}$. We stress here that although the order of computational complexity of one-bit GDA-Ada-1 and one-bit GDA-Ada-2 data detectors is the same as the Bayesian-based methods [13], [14] (see Table II), it will be illustrated in the next section that the run times needed for implementing one-bit GDA-Ada-1 and one-bit GDA-Ada-2 data detectors are dramatically lower than the run time needed for implementing the BiGAMP-based data detector (see Fig. 6). In addition, the computations for one-bit GDA-Ada-1 and one-bit GDA-Ada-2 data detectors can be straightforwardly parallelized.

## IV. SIMULATION RESULTS

In this section, numerical result that demonstrate the efficiency as well as superiority of the proposed wideband channel estimators and data detectors compared to other existing techniques are presented. In terms of computational complexity and run time, the AdaBoost-based methods are highly efficient, particularly when considering one-bit large-scale MIMO-OFDM systems. We use $T = 10$ for the proposed AdaBoost-based channel estimators and data detectors, unless otherwise stated. For channel estimation figures, orthogonal pilots are employed analogous to those suggested in [5, Eq. (23)]. In addition, quadrature phase shift keying (QPSK) constellations are used as the frequency domain symbols in data detection figures. The hyperparameter $C$ is set to 1 for SVM-based channel estimator and data detector of [17]. Furthermore, the modified finite Newton (MFN) method [38] is used for implementing the $\ell_2$-SVMs as it is one of the most efficient algorithms [17]. Performance of different channel estimators and data detectors are compared in terms of normalized MSE (NMSE) and bit-error-rate (BER), respectively. The former is defined as

$$\mathrm{NMSE} = \frac{\mathbb{E}\{\|\mathbf{H} - \hat{\mathbf{H}}\|_{\mathrm{F}}^2\}}{KM}$$

where $\mathbf{H} \triangleq [\mathbf{h}_1^{\mathrm{TD}}, \mathbf{h}_2^{\mathrm{TD}}, \ldots, \mathbf{h}_M^{\mathrm{TD}}]$ and $\hat{\mathbf{H}} \triangleq [\hat{\mathbf{h}}_1^{\mathrm{TD}}, \hat{\mathbf{h}}_2^{\mathrm{TD}}, \ldots, \hat{\mathbf{h}}_M^{\mathrm{TD}}]$. The block-fading interval is divided into two parts, where the first part and second part are used for channel estimation and data detection, respectively. Noteworthy to mention that the performance of one-bit GDA-Ada method (when the covariance matrix has to be computed exactly) is not reported in data detection figures as its computational complexity is higher than those of the proposed one-bit GDA-Ada-1 and one-bit GDA-Ada-2 methods for achieving similar performance.

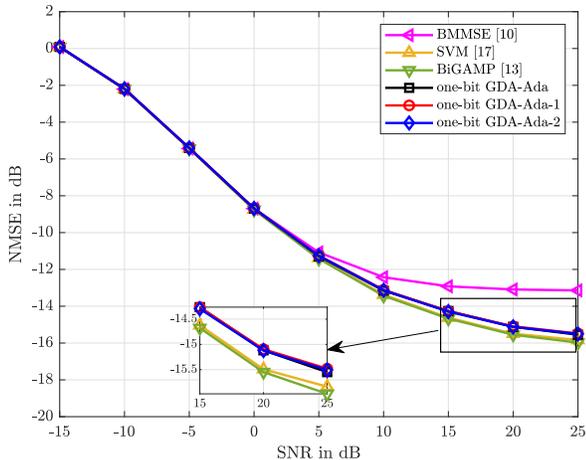

Fig. 1. Performance comparison of different channel estimators with $K = 2$, $M = 16$, $N_\mathrm{c} = 256$, and $L_\mathrm{tap} = 8$.

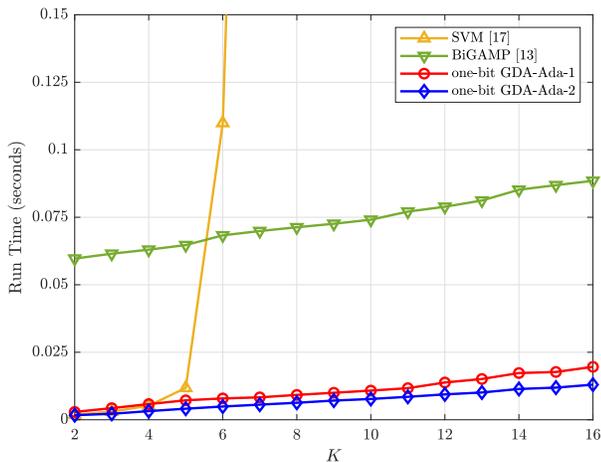

Fig. 2. Average run time comparison of the proposed one-bit GDA-Ada-1 and one-bit GDA-Ada-2 with SVM and BiGAMP in estimating channel between users and one antenna of the BS vs. the number of users $K$, considering the scenario where $N_\mathrm{c} = 512$, and $L_\mathrm{tap} = 16$.

In Fig. 1, the NMSE of the proposed AdaBoost-based channel estimators are compared with those of BMMSE [10], BiGAMP [13], and SVM [17]. It can be observed that the performance of BMMSE is worse than other methods tested, while BiGAMP possesses the best performance. The AdaBoost-based channel estimators are very marginally outperformed by BiGAMP and SVM that has no effect on the follow up data detection.

Despite having comparable channel estimation performance, the proposed one-bit GDA-Ada-1 and one-bit GDA-Ada-2 require substantially lower computational complexity compared to those of the SVM-based and BiGAMP-based methods as depicted in Fig. 2. We compare the required average run time for estimating channel between users and one antenna of the BS (i.e., average run time for estimating $\mathbf{h}_i^{\mathrm{TD}}$'s). Although the average run times for performing the channel estimation task are comparable for the one-bit GDA-Ada-1, one-bit GDA-Ada-2, and SVM-based methods when $K \leq 5$, the SVM-based channel estimator needs much higher computational complexity than those of the one-bit GDA-Ada-1 and one-bit GDA-Ada-2 when $K > 5$. In addition, the average run time for implementing the BiGAMP-based channel estimator is significantly larger for all $K$'s compared to the average run times required for executing the one-bit GDA-Ada-1 and one-bit GDA-Ada-2 channel estimators. We stress here that this advantage of the proposed methods is rooted in using low computation demanding techniques as weak classifiers in Algorithm 2. Moreover, the computational complexity order of different channel estimators tested is listed in Table I, where $\kappa(\cdot)$ represents a super-linear function and $I$ is the number of iterations required for implementing the BiGAMP method.

TABLE I: Order of Computational Complexity for Different Channel Estimators.

| Method | Complexity |
|---|---|
| **BMMSE** | $\mathcal{O}\left(M^2 K L_\mathrm{tab} N_\mathrm{c}\right)$ |
| **SVM-based** | $\mathcal{O}\left(MKL_\mathrm{tab} N_\mathrm{c} \kappa(N_\mathrm{c})\right)$ |
| **BiGAMP** | $\mathcal{O}\left(IMKL_\mathrm{tab} N_\mathrm{c}\right)$ |
| **one-bit GDA-Ada** | $\mathcal{O}\left(TM\max\{(KL_\mathrm{tab})^{2.373}, (KL_\mathrm{tab})^2 N_\mathrm{c}\}\right)$ |
| **one-bit GDA-Ada-1** | $\mathcal{O}\left(TMKL_\mathrm{tab} N_\mathrm{c}\right)$ |
| **one-bit GDA-Ada-2** | $\mathcal{O}\left(TMKL_\mathrm{tab} N_\mathrm{c}\right)$ |

In Fig. 3, a performance of the one-bit GDA-Ada-2 channel estimator is presented versus $T$ for SNR $\in \{-5, 5, 15, 25\}$ dB.[3] It can be seen that the channel estimation accuracy does not change substantially when $T > 10$ for SNR $\in \{-5, 5, 15, 25\}$ dB. As a result, opting $T = 10$ in Algorithm 2 is a reasonable choice according to Fig. 3.

Fig. 4 compares the NMSE of the proposed AdaBoost-based channel estimators with SVM and BiGAMP for $N_\mathrm{c} = 256$ and $N_\mathrm{c} = 1024$, where the NMSEs of the methods tested are decreased for about $4$ dB at high SNRs by increasing $N_\mathrm{c}$ from 256 to 1024. Analogous to Fig. 1, the proposed AdaBoost-based channel estimators possess quite similar performance to the performance of the SVM-based and BiGAMP-based channel estimators.

Fig. 5 compares the one-bit GDA-Ada-1 and one-bit GDA-Ada-2 data detectors with the SVM and BiGAMP data detectors for both cases of estimated CSI and perfect CSI. It should be noted here that the estimated CSI of each method is found by their corresponding channel estimators. Moreover, 500 independently generated CSIs are considered for calculating the BERs. It can be seen in Fig. 5 that the BERs of the proposed AdaBoost-based data detectors outperform the BERs of the SVM and BiGAMP data detector at high SNRs for both cases of estimated CSI and perfect CSI. In addition, Fig. 5 shows that the performance of the one-bit GDA-Ada-1 method is slightly better than that of the one-bit GDA-Ada-2 method at high SNRs. The better performance of the one-bit GDA-Ada-1 and one-bit GDA-Ada-2 methods indicates that the proposed methods are more robust to/independent of specific channel realizations than the the SVM and BiGAMP-based methods. Indeed, the performance of the BiGAMP method, for example, shows sensitivity to channel realizations because it saturates and worsens compared to the performance of the one-bit GDA-Ada-1 and one-bit GDA-Ada-2 methods at high SNR, meaning that for some channel realizations the performance of the BiGAMP method may be significantly worse than that of the one-bit GDA-Ada-1 and one-bit GDA-Ada-2 methods.

---

[3]As the behavior of all proposed Adaboost-based channel estimators with respect to $T$ follows the same pattern, only the performance of the one-bit GDA-Ada-2 is shown in Fig. 3 to ensure the clarity of presentation.

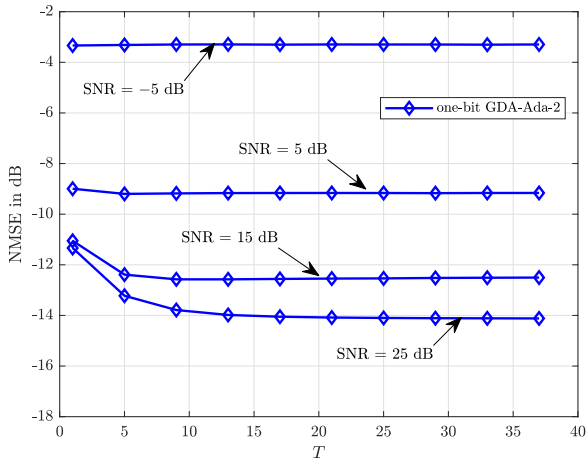

Fig. 3. NMSE comparison of the one-bit GDA-Ada-2 channel estimators for different values of $T$ with $K = 4$, $M = 32$, $N_c = 512$, and $L_{\text{tap}} = 16$.

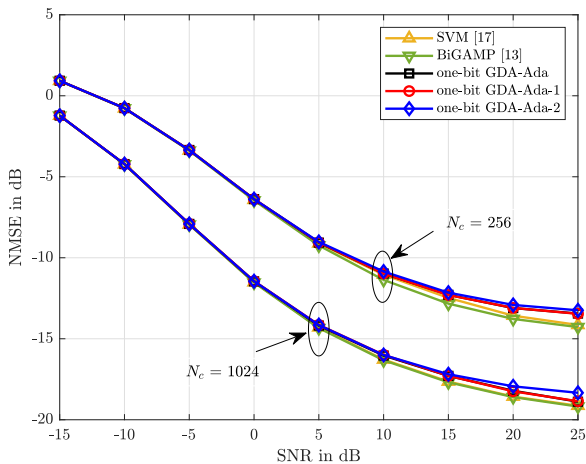

Fig. 4. Performance comparison between the proposed AdaBoost-based channel estimators, SVM, and BiGAMP with $K = 2$, $M = 32$, $L_{\text{tap}} = 16$, and $N_c \in \{256, 1024\}$.

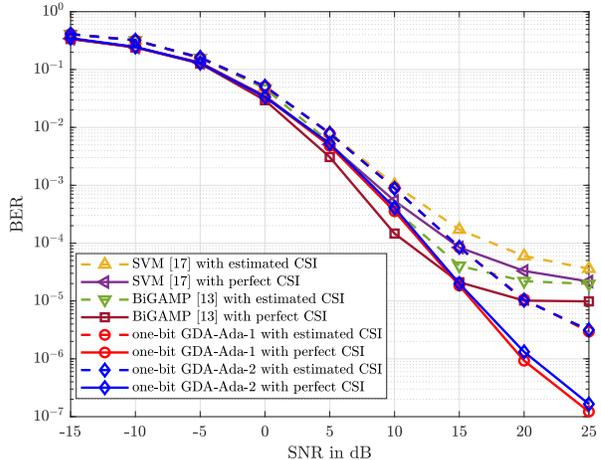

Fig. 5. Performance comparison of different data detectors with $K = 2$, $M = 16$, $N_c = 256$, $L_{\text{tap}} = 8$, and QPSK modulation.

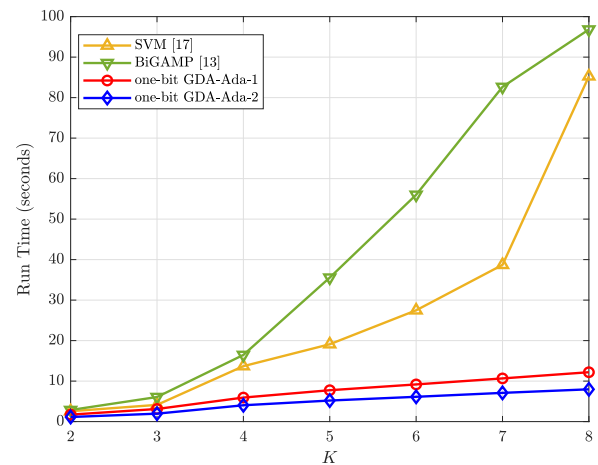

Fig. 6. Average run time for implementing different data detectors with various $K$, $N_c = 512$, $M = 32$, and $L_{\text{tap}} = 8$.

An average run time comparison for implementing the one-bit GDA-Ada-1, one-bit GDA-Ada-2, SVM, and BiGAMP data detectors is presented in Fig. 6, where $K \in \{2, 3, \ldots, 8\}$. For $K \geq 4$, the average run times of the SVM and BiGAMP data detectors are substantially higher than those of the one-bit GDA-Ada-1 and one-bit GDA-Ada-2 data detectors. The orders of computational complexity of different data detectors tested are listed in Table II.

The impact of choosing different $T$ on the performance of the one-bit GDA-Ada-2 data detector for SNR $\in \{-5, 0, 10, 20\}$ dB is investigated in Fig. 7.[4] It can be seen that the change in BER is only marginal when $T > 10$ for SNR $\in \{-5, 0, 10, 20\}$ dB. As a result, opting $T = 10$ in Algorithm 3 is a reasonable choice according to Fig. 7.

---

[4]As the behavior of all proposed Adaboost-based data detectors with respect to $T$ follows the same pattern, only the performance of the one-bit GDA-Ada-2 is shown in Fig. 7 to ensure the clarity of presentation.

TABLE II: Order of Computational Complexity for Different Data Detectors.

| Method | Complexity |
|--------|-----------|
| **SVM-based** | $\mathcal{O}\left(MKN_c^2\kappa(MN_c)\right)$ |
| **BiGAMP** | $\mathcal{O}\left(IMKN_c^2\right)$ |
| **one-bit GDA-Ada** | $\mathcal{O}\left(T\max\{(KN_c)^{2.373}, MK^2N_c^3\}\right)$ |
| **one-bit GDA-Ada-1** | $\mathcal{O}\left(TMKN_c^2\right)$ |
| **one-bit GDA-Ada-2** | $\mathcal{O}\left(TMKN_c^2\right)$ |

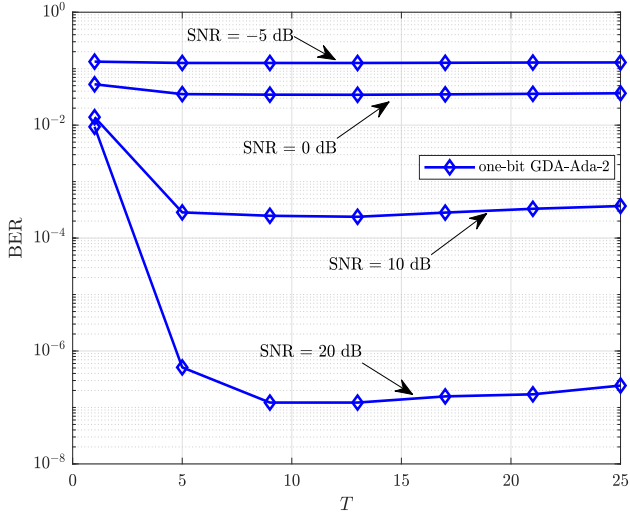

Fig. 7. The impact of different $T$ on the performance of the one-bit GDA-Ada-2 data detector with $K = 4$, $N_c = 256$, $M = 32$, $L_{tap} = 16$, and QPSK modulation.

## V. CONCLUSION

In this paper, we have found out and demonstrated that the GDA classifier/approximated GDA classifier together with the AdaBoost technique result in developing efficient and reliable channel estimators and data detectors, specifically in large scale scenarios such as MIMO-OFDM systems that operate over frequency selective channels. It was shown that two of the proposed AdaBoost-based channel estimators and data detectors named one-bit GDA-Ada-1 and one-bit GDA-Ada-2 require dramatically lower run time compared to those of the BiGAMP-based and SVM-based methods, while providing comparable/better accuracy. Numerical results were presented to showcase the efficiency and robustness of the proposed methods in large scale MIMO-OFDM systems. For one-bit MIMO-OFDM systems, the use of AdaBoost with weak classifiers can be viewed as a versatile framework where any approximated binary classifiers with low computational complexity can be employed as weak classifiers, resulting in this AdaBoost framework being a highly promising tool for dramatically reducing the computational complexity.